\pdfoutput=1
\documentclass[final,11pt]{elsarticle}
\usepackage[top=2.0cm,bottom=2.0cm,left=2.0cm,right=2.0cm]{geometry}
\usepackage{amssymb}
\usepackage{graphicx}
\usepackage{subfigure}
\usepackage{caption}
\usepackage{hyperref}
\usepackage{siunitx}
\usepackage{psfrag}
\usepackage{pstool}
\usepackage{changes}
\usepackage{commath}
\usepackage[normalem]{ulem}
\usepackage[numbers]{natbib}
\usepackage{floatrow}
\usepackage{placeins}
\usepackage[symbol]{footmisc}
\usepackage{bm}
\usepackage{amsmath}
\usepackage{chngcntr}
\usepackage{color,soul}
\usepackage{lineno}
\usepackage{multirow}

\journal{Journal of Materials Science \& Technology}

\begin{document}

\begin{frontmatter}

\title{Macroscopic analysis of time dependent plasticity in Ti alloys}

\def\correspondingauthor{\footnote{*Corresponding author.}}
\author[add1]{Yi Xiong\corref{*}}
\cortext[*]{Corresponding author}
\address[add1]{Department of Materials, University of Oxford, Parks Road, Oxford, OX1 3PH, United Kingdom}
\ead{yi.xiong@materials.ox.ac.uk}

\author[add1]{Phani S. Karamched}

\author[add2]{Chi-Toan Nguyen}
\address[add2]{Safran SA, Safran Tech, Department of Materials and Processes, 78772 Many-les-Hameaux, France}

\author[add3]{David M. Collins}
\address[add3]{School of Metallurgy and Materials, University of Birmingham, Edgbaston, Birmingham, B15 2TT, United Kingdom}

\address[add4]{Department of Engineering Science, University of Oxford, Parks Road, Oxford, OX1 3PJ, United Kingdom}
\author[add1]{Christopher M. Magazzeni}
\author[add4,add1]{Edmund Tarleton}
\author[add1]{Angus J. Wilkinson}

\begin{abstract}

Component failure due to cold dwell fatigue of titanium and its alloys is a long-standing problem which has significant safety and economic implications to the aviation industry. This can be addressed by understanding the governing mechanisms of time dependent plasticity behaviour of Ti at low temperatures. Here, stress relaxation tests were performed at four different temperatures on three major alloy systems: commercially pure titanium (two alloys with different oxygen content), Ti-6Al-4V (two microstructures with differing  $\beta$ phase fractions) and Ti-6Al-2Sn-4Zr-$x$Mo (two alloys with different Mo content $x$=2 or 6, and portion of $\beta$ phase). Key parameters controlling the time dependent plasticity were determined as a function of temperature. Both activation volume and energy were found to increase with temperature in all six alloys. It was found that the dwell fatigue effect is more significant by oxygen alloying but is suppressed by the addition of Mo. The presence of the $\beta$ phase did not strongly affect the dwell fatigue, however, it was suppressed at high temperature due to the low strain rate and strain rate sensitivity.   
\end{abstract}


\end{frontmatter}



Cold dwell fatigue is the critical technological issue in Ti alloys used in the cooler front sections of aerospace jet engines. When subjected to a sustained load (stress dwell), the lifetime of these Ti components can be drastically reduced by as much as an order of magnitude~\cite{BACHE2003,DUNNE2008,ZHENG201743,NEERAJ2000}.  Without a clear understanding of the effect, safety concerns are raised. Aeroengine manufacturers currently manage this problem through heavily conservative component design and operation. Cold dwell fatigue is mainly manifested in the $\alpha$ phase (HCP) Ti~\cite{CONRAD1981,CONRAD2011}, due to its anisotropic property~\cite{LEI202177,ZHANG2021265}. Computational models indicate that crack nucleation during cold dwell fatigue is dominated by so-called rogue hard-soft grain pairs. During a stress dwell period, localised time dependent plasticity in the ‘soft’ grain (grain oriented favourably for the dominate slip systems) leads to load shedding and an increase in elastic stress within the neighbouring ‘hard’ grain (grain oriented badly for the dominant slip systems)~\cite{DUNNE20071061,HASIJA2003}. With repeated load cycling, the stress in the ‘hard’ grain may become sufficiently high to initiate cracking, which ultimately leads to catostraphic failure.

Commonly used alloys in aeroengine compressor disks, such as Ti-6Al-2Sn-4Zr-$x$Mo alloys show accelerated failure via the accumulation of plastic strains when subjected to a dwell stress~\cite{Qiu2014}. The role of bulk composition is known to influence dwell damage; the propensity for dwell debit is minor in high-Mo containing Ti6246 alloys, whereas it is significant in low-Mo containing Ti6242 alloys which have a lower $\beta$ phase fraction, and is relatively minor in Ti6246 with a greater $\beta$ phase fraction. The role of chemistry and microstructure, both known to affect the severity of dwell fatigue, is not well established. Although this problem has drawn attention for many decades and great efforts have been made to understand this problem, the majority of the work has been performed at room temperature. In practice, it is found that cold dwell fatigue is suppressed at elevated temperatures, and is no longer present for operating temperatures exceeding 200~$^{\circ}$C~\cite{TITANIUM,ZHANG2015}. To establish a better understanding of the cold dwell fatigue problem, a broader study that explicitly explores the sensitivity of temperature to time dependent plasticity is necessary.  

This proposed experiment seeks to gain mechanistic understanding and quantification of key parameters across a range of industrially relevant Ti alloys and microstructures by using simple stress relaxation loading at different temperatures up to 250~$^{\circ}$C. The effect of chemistry, microstructure and temperature on time dependent plasticity has been explored here, which aims to benefit the future design of alloys and provide better lifetime prediction under conditions observed in service. Although not capturing the local extreme conditions at the grain level, the macroscopic stress-strain-time response gives readily accessible route to assess the propensity for different alloy compositions, microstructures and textures to influence the extent of time dependent plasticity in the temperature regime of interest for cold dwell behaviour. Macroscopic assessment of time dependent plasticity is the aim of this paper.

\begin{table*}[t]
\newcommand{\tabincell}[2]{\begin{tabular}{@{}#1@{}}#2\end{tabular}} 
\footnotesize
\centering
\caption{Materials composition (in wt.\%) and microstructure information.}
 \begin{tabular*}{1\textwidth}{@{\extracolsep{\fill}}c|@{\extracolsep{\fill}}c@{\extracolsep{\fill}}c@{\extracolsep{\fill}}c@{\extracolsep{\fill}}c@{\extracolsep{\fill}}c@{\extracolsep{\fill}}c@{\extracolsep{\fill}}c@{\extracolsep{\fill}}c@{\extracolsep{\fill}}c@{\extracolsep{\fill}}c@{\extracolsep{\fill}}c@{\extracolsep{\fill}}c} 
 \hline
  & Al & V & Sn & Zr & Mo & Fe & O & N & \tabincell{c}{Others\\ total} & Microstructure & \tabincell{c}{$\beta$ phase\\content} &  \tabincell{c}{Grain size\\($\mu${m})}\\
 \hline
 CP-Ti grade 1 & $<$0.05 & 0.11 & $<$0.05 & $<$0.05 & $<$0.05 & $<$0.05 & 0.077 & 0.02 & 0.2 & Pure $\alpha$ & 0\% & 30 \\ 

 CP-Ti grade 4 & $<$0.05 & 0.15 & $<$0.05 & $<$0.05 & $<$0.05 & 0.055 & 0.32 & 0.006 & 0.4 & Pure $\alpha$ & 0\% & 17 \\

 Ti64 $\alpha$ & 6.74 & 3.98 & $<$0.05 & $<$0.05 & $<$0.05 & $<$0.05 & 0.218 & 0.001 & 0.2 & Primary $\alpha$ & 6\% & 15 \\

 Ti64 Bimodal & 6.63 & 3.87 & $<$0.05 & $<$0.05 & $<$0.05 & $<$0.05 & 0.191 & 0.002 & 0.3 & $\alpha+\beta$ & 18\% & 8 \\

 Ti6242 & 6.21 & 0.28 & 1.27 & 4.21 & 1.86 & $<$0.05 & 0.079 & $<$0.001 & 0.4 & Primary $\alpha$ & 10\% & 12 \\

 Ti6246 & 5.98 & 0.15 & 0.87 & 4.24 & 5.35 & $<$0.05 & 0.157 &$<$0.001 & 0.3 & $\alpha+\beta$  & 60\% & 6 \\
 \hline
\end{tabular*}
\label{table.1}
\end{table*}

\begin{figure*}[htbp]
\centering
{
\includegraphics[width=0.85\textwidth]{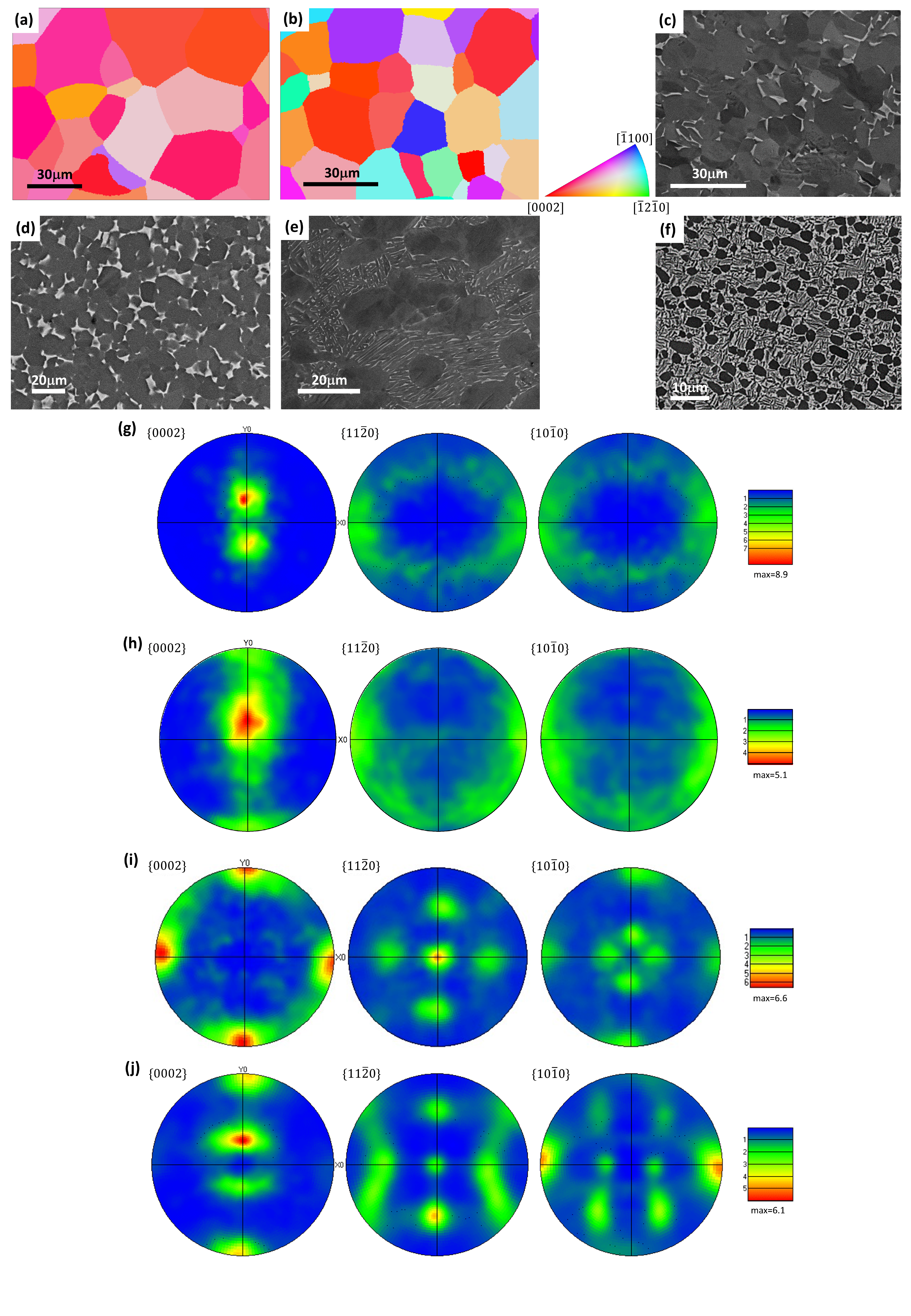}
}
\caption{Microstructures of the raw materials: (a) EBSD map of the CP-Ti grade 1 alloy (IPF map in normal direction of the raw material); EBSD map of the CP-Ti grade 4 alloy (IPF map in normal direction of the raw material); SE images of the (c) Ti64 $\alpha$ alloy; (d) Ti64 Bimodal alloy; (e) Ti6242 alloy and (f) Ti6246 alloy; Pole figures of the raw material with a 10$^{\circ}$ width used for contours plotting calculation: (g) CP-Ti grade 1 alloy; (h) CP-Ti grade 4 alloy; (i) Ti64 $\alpha$ alloy and (j) Ti6242 alloy. }
\label{fig:1}
\end{figure*}

In this work, we have studied the stress relaxation behaviour of 6 different Ti alloys. An overview of the chemistry and microstructures of the alloys is summarised in Table~\ref{table.1}. Samples were cut from the raw materials then ground and polished to a colloidal silica finish. Microstructural analyses were conducted on these polished samples using a Zeiss Merlin scanning electron microscope (SEM) equipped with a Bruker e-flash detector, operating with a beam current of 20~nA and an accelerating voltage of 20~kV. Exemplar microstructures of these alloys are shown in Figure~\ref{fig:1}(a)-(f). The two grades of commercially pure Ti (CP-Ti) were used to study the effect of the common interstitial, oxygen, on cold dwell fatigue. Two Ti64 alloys with different microstructures were studied to establish the effect of the $\beta$ phase content. In the two Ti64 alloys, the $\beta$ phase was found to appear at the $\alpha$ grains boundaries and mainly at the triple junctions. The Ti6242 and Ti6246 alloys were used to study the effect of both the $\beta$ phase and the alloying element Mo on time dependent plasticity. In the Ti-6Al-2Sn-4Zr-$x$Mo alloys, the $\beta$ phase had a lamellar structure. Texture is an important factor that affects the deformation behaviour of $\alpha$-Ti due to the anisotropic properties of the HCP crystal structure. The CP-Ti grade 4, CP-Ti grade 1, Ti64 $\alpha$ and Ti6242, each supplied as rolled sheets, all have a volume fraction of the $\alpha$ phase exceeding 90\%. Their prior processing gives a strong texture in these four materials (as shown in Figure~\ref{fig:1}(g)-(j)). The pole figures, calculated from EBSD measurements, were obtained from larger areas that contains over 3000 grains. To ensure the major slip systems are activated, there should be a considerable number of ‘soft’ grains with respect to the loading direction. The test samples were cut from the raw materials with the loading direction along the X0 direction in these pole figures. 

\begin{figure*}[hbt!]
\centering
{
\includegraphics[width=0.8\textwidth]{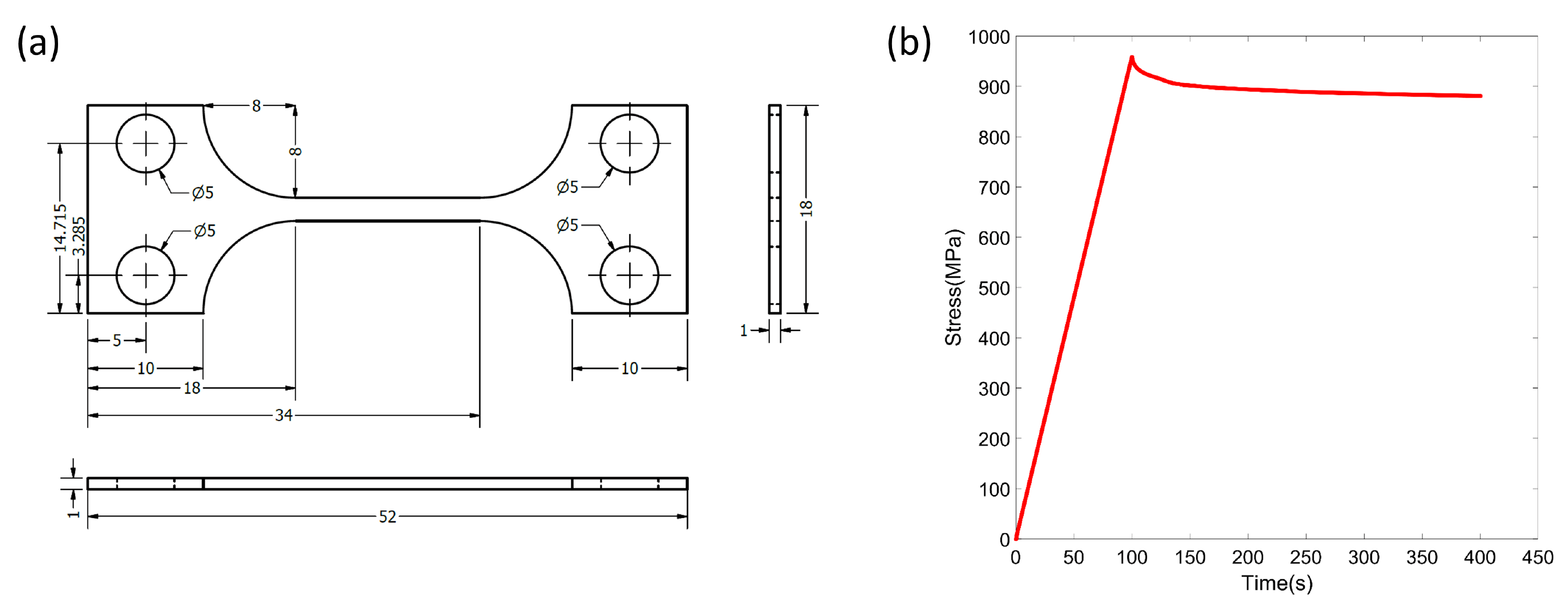}
}
\caption{(a) Geometry of the ETMT tensile test sample; (b) An example of the stress relaxation loading (Ti64 $\alpha$ at room temperature).}
\label{fig:2}
\end{figure*}

A series of samples in each alloy derivative were subjected to stress relaxation loading (load-up and strain hold) using an Instron ETMT (Electro-Thermal-Mechanical Tester), where test samples were subjected to user defined mechanical and thermal profiles. Sample temperature control was achieved through a DC electric current that was fixed between water cooling grips. R-type thermocouples were spot-welded to the centre of each sample gauge to provide accurate temperature surveillance during the tests. Sample deformation was applied through an in-line drive, displacement controlled 5 kN mechanical loading assembly~\cite{TANG2021116468,Roebuck2001,SULZER2018}. Figure~\ref{fig:2}(a) shows the geometry of the ETMT dogbone sample, having a 52~mm length and the gauge length of 16~mm with a 2~mm width and a 1~mm thickness. Each alloy was tested at four different temperatures: room temperature, 75$^{\circ}$C, 145$^{\circ}$C and 250$^{\circ}$C (CP-Ti grade 1 at 250$^{\circ}$C test was abandoned due to its extremely low strength at this temperature). A set of trial tests were performed to determine the target loads, where the macroscopic yield points were reached and were exceeded by a small plastic strain. This was assessed via changes in the gradient of the load-stroke (stress-strain) curves. Samples were then loaded up to the target loads at constant stress rates (different for different samples) in around 100 to 150s, corresponding to macroscopic strain rates of approximately $3\times10^{-5}~\textrm{s}^{-1}$ (as measured from the ETMT stroke rate) within the elastic region. A typical stress relaxation loading curve is shown in Figure~\ref{fig:2}(b).   

The stress relaxation curves were numerically analysed using a constitutive law~\cite{DUNNE20071061,XIONG2020,creep_analysis_constitutive,Kalyanasundaram2016,WANG2016656}, which links the macroscopic plastic strain rate, $\dot{\varepsilon}_p$, when $\sigma\geqslant\sigma_c$, to the macroscopic applied stress, $\sigma$:

\begin{equation}\label{eq:1}
    \dot{\varepsilon}_p = \rho b^2 \nu \exp\left(-\frac{\Delta{F}}{k_BT}\right)\sinh\left(\frac{(\sigma-\sigma_c) {\Delta}V}{k_BT}\right)
\end{equation}
\begin{gather}
   \nonumber \Delta{\varepsilon}_e = -\Delta{\varepsilon}_p = -\dot{\varepsilon}_p\Delta{t} \\
  \label{eq:2}  \rightarrow \Delta{\sigma} = E\Delta{\varepsilon_e} \\  \nonumber
    \rightarrow \sigma_{i+1} = \sigma_i + \Delta{\sigma}  
\end{gather}

\noindent where $\rho$ is the density of gliding dislocations, $b$ is the magnitude of the Burgers vector, $\nu$ is the jump frequency, $k_B$ is the Boltzmann constant. These parameters are constant and are summarised in Table~\ref{table:2}. $T$ is the absolute temperature. The remaining three terms are material dependent and are the key parameters controlling plasticity: $\Delta{F}$ is the thermal activation energy, $\sigma_{c}$ is the critical stress and $\Delta{V}$ is the activation volume~\cite{DUNNE20071061}. The three key parameters were determined via this constitutive law and the experimental stress relaxation curves.  

The mobile dislocation density, $\rho$, in Equation~(\ref{eq:1}) was considered as an averaged value and was held constant in this equation (the value of $\rho$ was taken from other studies on CP-Ti, Ti6Al, Ti6242 and Ti6246 alloys on cold dwell fatigue)~\cite{ZHANG2016Intrinsic,ZHANG2017199,ASHTON2017377,XIONG2021116937}. The validation of this assumption can be found in prior work~\cite{XIONG2021116937}. The sensitivity of the three fitted parameters ($\Delta{V}$, $\Delta{F}$ and $\sigma_c$) to change of $\rho$  can be found in~\ref{sec:app}.

\begin{table}[h]
\begin{tabular*}{0.35\textwidth}{c@{\extracolsep{\fill}}c}
\hline
$\rho$ & 5~$\mu\text{m}^{-2}$ \\
$b$ & 0.295~nm \\
$\nu$ & $10^{11}$~Hz  \\
$k_B$ &  $1.38 \times 10^{-23}~\text{JK}^{-1}$\\
\hline
\end{tabular*}
\caption{Values of the fixed parameters in the slip law~\cite{CUDDIHY2017,ZHANG2016Intrinsic,ASHTON2017377,ZHANG2017199,XIONG2021116937}}
\label{table:2}
\end{table}

\begin{figure*}[htbp]
    \centering
    \includegraphics[width=1\textwidth]{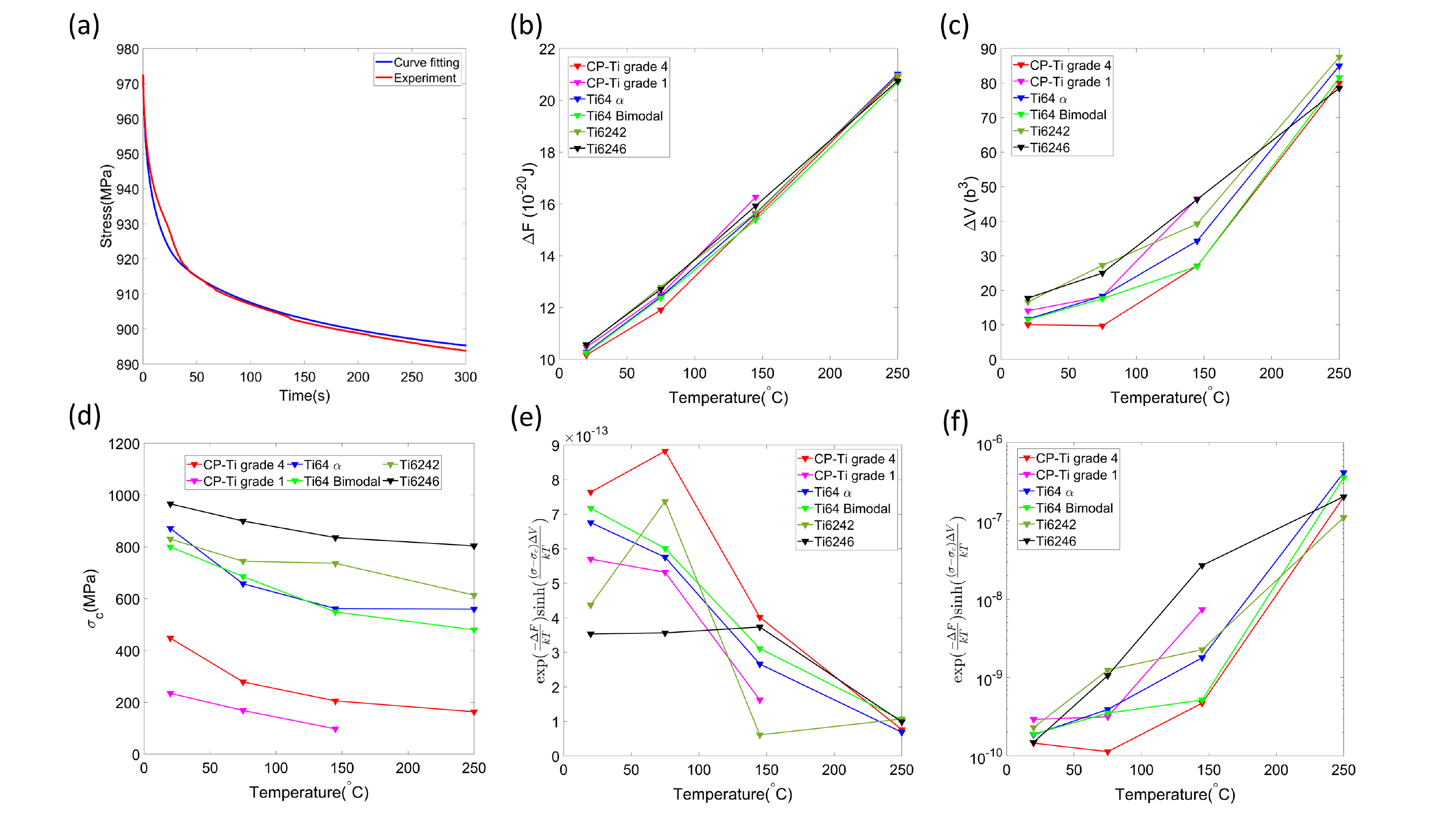}
    \caption{(a) A fitted stress relaxation curve (Ti64 $\alpha$ at room temperature, as an example). (b) Thermal activation energy, $\Delta{F}$ and  (c) Activation volume, $\Delta{V}$ (d) Critical stress, $\sigma_c$, as a function of temperature for the 6 alloys; (e) Contribution of the three key parameters to the form of the plastic strain rate when the critical stress is exceeded by 1~MPa (low stress condition) and (f) 50~MPa (high stress condition);}
    \label{fig:3}
\end{figure*}

To obtain the slip law parameters, the following steps were used, as shown in Equation~(\ref{eq:2}), the initial applied stress, $\sigma_i$, was substituted into the Equation~(\ref{eq:1}) to generate the plastic strain rate, $\dot{\varepsilon}_p$. The change in plastic strain, $\Delta{\varepsilon}_p$, over a time step, $\Delta{t}$ (here taken to be 0.2~s) was then be calculated by the product of $\dot{\varepsilon}_p$ and $\Delta{t}$. The change in elastic strain, $\Delta{\varepsilon}_e$, was matched by the negative change in plastic strain, $\Delta{\varepsilon}_p$ during the stress relaxation (as the total strain was held constant). The change in stress, $\Delta{\sigma}$, was obtained by multiplying the Young's modulus, $E$ for Ti at each temperatures~\cite{ZHANG2015,STRINGER1960} to the elastic strain, $\Delta{\varepsilon}_e$. The stress at the next time step, $\sigma_{i+1}$,was then calculated by adding the $\Delta{\sigma}$ to the $\sigma_i$. 
This process was repeated so that a modelled stress relaxation curve could be generated. This method is widely used for creep stress analysis~\cite{boyle2013stress}. Three key parameters were optimised using a fitting tool in MATLAB so that the modelled relaxation curves and the experimental relaxation curves show good agreement with errors of less than 1\%, as shown in Figure~\ref{fig:3}(a). The confidence intervals for the three parameters are $\pm0.2\times10^{-20}$J for $\Delta{F}$, $\pm1b^3$ for $\Delta{V}$ and $\pm5$MPa for $\sigma_c$. This numerically fitted parameters are broadly similar to those used by other studies of titanium alloys~\cite{ZHENG201743,DUNNE20071061,ZHANG2016Intrinsic}.

Figure~\ref{fig:3}(b)-(d) show the three key parameters vs. temperature for the six different alloys. Both $\Delta{F}$ and $\Delta{V}$ increased with temperature, where the relationship between $\Delta{F}$ and temperature was almost linear, which is consistent with the observations in CP-Ti and BCC steel~\cite{TANAKA2016,XIONG2021116937}. The effect of temperature on $\Delta{F}$ was seen to be markedly stronger than the alloy-to-alloy variation. CP-Ti had the lowest critical stress and showed a comparably lower $\Delta{V}$ among the three different alloy systems. Comparing the two CP-Ti alloys, the grade 4 with a higher oxygen content had a higher critical stress over the low oxygen content grade 1, corroborating the known potent strengthening effect of oxygen in Ti~\cite{Yu2015,BRITTON2015}. A higher activation volume was found in the low oxygen content grade 1 samples over the grade 4. The Ti-6Al-4V alloys system had both moderate critical stress and activation volume magnitudes among the three alloy systems. With the addition of Al and V into the pure Ti, the strength was largely increased (i.e. the $\sigma_c$ of Ti64 $\alpha$ had approximately double the value for the CP-Ti grade 4 and triple the value for the grade 1). With the presence of $\beta$ phase in Ti64, $\Delta{V}$ was slightly reduced, especially at higher temperatures (145~$^{\circ}$C and 250~$^{\circ}$C), but $\sigma_c$ was not significantly affected. With further alloying additions, the most complex alloys system Ti-6Al-2Sn-4Zr-$x$Mo system had a relatively higher $\Delta{V}$ and the highest $\sigma_c$. There is no significant difference in $\sigma_c$ between the Ti6242 and the two Ti64 alloys, but $\sigma_c$ for the Ti6246 is evidently higher than them. This shows the excellent strengthening effect of  Mo~\cite{CORREA2015180}, where its presence acts as a grain refiner (recall Table~\ref{table.1} the Ti6246 has the finest grain size) and further improves the mechanical strength by the Hall–Petch effect. The higher Mo content Ti6246 had a higher $\sigma_c$ over the Ti6242, but there was no obvious trend between the $\Delta{V}$ of Ti6242 and Ti6246.  

The activation volume, $\Delta{V}$, expresses the volume which is physically swept by a dislocation from a ground equilibrium state to an activated state after the deformation, and it is proportional to the dislocation pinning distance (or the mean free path of dislocation)~\cite{DUNNE20071061,Sarkar2015}. In CP-Ti, interstitial oxygen atoms provide a strengthening effect by acting as obstacles for dislocations. The pinning distance reduces when there are more oxygen atoms, resulting in a smaller activation volume in the higher oxygen content CP-Ti grade 4 compared to the CP-Ti grade 1. Alloying is likely to increase the pinning distance, because the Ti-6Al-4V alloys have a slightly higher $\Delta{V}$ over the CP-Ti alloys, and with further alloying, the Ti-6Al-2Sn-4Zr-$x$Mo system has the highest $\Delta{V}$ among the three alloy systems. Mo, as a substitutional elements in Ti, was found to increase the lattice spacing~\cite{Fellah2019}, which is likely to lead to a higher dislocation pinning distance. The $\Delta{V}$ magnitudes increase with temperature for all six alloys; alloying has no effect on this trend. Therefore, this trend could solely depend on  thermal vibrations. Stronger thermal vibrations at higher temperatures could lead to a larger dislocation pinning distance and thus a higher activation volume. Grain boundaries are also likely to affect the dislocation process, and thus the $\Delta{V}$ could be affected by the grain size of the samples. However, this grain size effect becomes important only when the grain size is reduced to below 30-50 nm~\cite{WANG20062715,CHENG20034505}. As all the Ti samples examined in this work have grain sizes at the micrometer level, the effect of grain size on a dislocation process or the activation volume was not considered. The value of $\Delta{V}$ is an effective reflection of rate-dependent deformation mechanisms~\cite{Sarkar2015}. The intersection of forest dislocations usually gives rise to an extremely large activation volume (several thousand $b^3$). A dislocation climb process, however, yields a lower activation volume of 1$b^3$~\cite{Sarkar2015}. Overcoming the Peierls barrier, point defect drag and point defect interactions typically give an activation volume of 10$b^3$ to 100$b^3$ according to Evans and Rowlings~\cite{Evans_and_Rawlings}. In this work, the values of $\Delta{V}$ of all six alloys at all four temperatures were found to lie within this range, indicating the rate-dependent deformation mechanism remains unchanged and the mechanism is not affected by alloying or temperatures.       

Figure~\ref{fig:3}(e) and (f) show the contribution of the three key parameters into the plastic strain rate and thus time dependent plasticity. The prefactor $\rho{b}^2\nu=4.35\times10^{4}s^{-1}$ is constant in Equation~(\ref{eq:1}) and was ignored in this calculation. The contribution of the three key parameters to the plastic strain rate was calculated when $(\sigma-\sigma_c)$=1~MPa (low stress condition, see Figure~\ref{fig:3}(e)) and when $(\sigma-\sigma_c)$=50~MPa (high stress condition, see Figure~\ref{fig:3}(f)). When the critical stress was just exceeded, similar to the dwell fatigue condition, the contribution decreased as the temperature increased from room temperature to 250$^{\circ}$C and the values were still of the same order of magnitude. Generally, CP-Ti grade 4 had the highest strain rate, Ti-6Al-4V systems had a moderate strain rate and Ti6246 had the lowest strain rate. All six alloys had a very low strain rate at 250$^{\circ}$C, and in most of the alloys, there was a sharp drop in strain rate when the temperature reached 145$^{\circ}$C. For the high stress condition, the trend is reversed, where the contribution of key parameters to the plastic strain rate increased with temperature and the difference could be up to 4 orders of magnitude. The CP-Ti grade 4 had the lowest strain rate and Ti6246 had the highest. In both stress conditions, the two Ti64 alloys show relatively similar strain rates. The deformation behaviour was not strongly affected by the $\beta$ phase fraction.

\begin{figure*}[hbt!]
    \centering
    \includegraphics[width=0.9\textwidth]{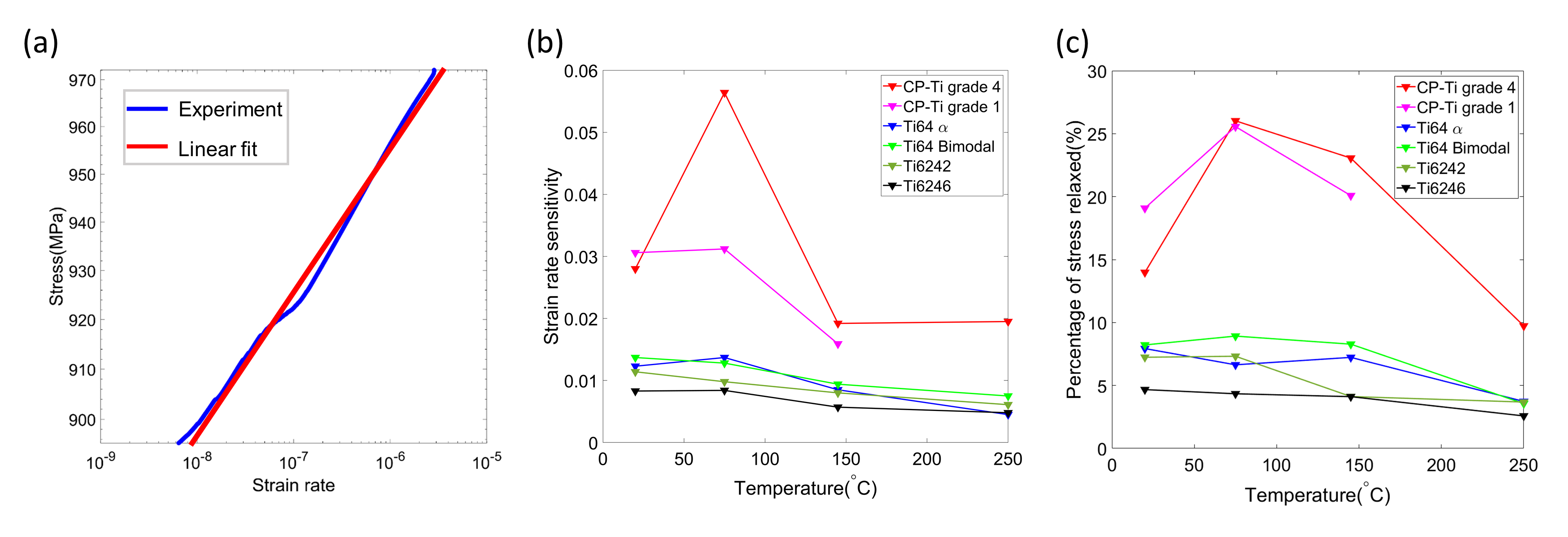}
    \caption{(a) Fitting a straight line to the stress vs. strain rate on a log scale, the gradient represents the strain rate sensitivity exponent $m$ (Ti64 $\alpha$ at room temperature as an example); (b) Strain rate sensitivity vs. Temperatures for the six different alloys; (c) Percentage of stress relaxed vs. Temperatures for the six different alloys.}
    \label{fig:4}
\end{figure*}

The strain rate sensitivity (SRS) exponent, $m$, can be calculated follow the equation~\cite{JUN2016,JUN2016NANO}:
\begin{equation}\label{eq11}
    m = \rm{d}(\log({\sigma}))/\rm{d}(log({\dot{\varepsilon}}))
\end{equation}
In this stress relaxation experiment, the $m$ values are determined by fitting a straight line to the stress-strain rate plots on a log scale. The gradient of the straight line is the strain rate sensitivity exponent; more details of this method can be found in previous work~\cite{XIONG2020,XIONG2021116937}. As shown in Figure~\ref{fig:4}(b), the commercially pure Ti had the highest $m$ values, Ti-6Al-4V had a medium SRS and the values for the Ti-6Al-2Sn-4Zr-$x$Mo system was the lowest. The SRS for CP-Ti grade 4 was the highest and Ti6246 was the lowest among the six alloys. The percentage of stress relaxation was calculated to represent the accumulation of plastic strain during a stress relaxation cycle, as the total strain was held constant. The relaxed stress (elastic strain) was matched by the increased plastic strain. A higher strain rate sensitivity will lead to more plastic deformation and thus higher plastic strain during stress relaxation~\cite{XIONG2020,XIONG2021116937}, as a result, the trend found in Figure~\ref{fig:4}(c) is very similar to the trend in Figure~\ref{fig:4}(b). It is also noted that, there were sharp increases in SRS and relaxed stress as temperature increases from room temperature to 75$^{\circ}$C in the two CP-Ti alloys. This indicates the strong dislocation activity at this intermediate temperature. This temperature agrees well with the predicted worst-case scenario temperature (90 to 120~$^{\circ}$C) for dwell debit according to Zhang \textit{et al.}~\cite{ZHANG2015}.        

To summarise, stress relaxation tests were performed on six different alloys at 4 different temperatures up to 250$^{\circ}$C. Key parameters (thermal activation energy $\Delta{F}$, activation volume $\Delta{V}$ and critical stress $\sigma_c$) governing the time dependent plasticity were obtained by fitting the macroscopic stress relaxation cycles to the slip law. Commercially pure Ti has the lowest strength but the highest strain rate sensitivity, as a result, the dwell fatigue effect is the strongest among the three alloy systems. Ti-6Al-4V has both moderate strength and strain rate sensitivity, the dwell fatigue effect was not significantly influenced by the fraction of $\beta$ phase in Ti64. The Ti-6Al-2Sn-4Zr-$x$Mo system had the highest strength but the lowest strain rate sensitivity, dwell fatigue effect was highly suppressed in these alloys. Oxygen has an excellent strengthening effect but does enhance the dwell fatigue effect. Mo evidently offers both a strengthening effect and significant resistant to dwell fatigue. At dwell fatigue stress conditions (when critical stress is just exceeded), both the contribution of the key parameters to strain rate and strain rate sensitivity sharply decreased in all six alloys when temperature is above 145$^{\circ}$C. As a result, plastic strain accumulation at 250$^{\circ}$C is at a very low level, and dwell fatigue is suppressed at high temperature. Furthermore, the authors believe that the cold dwell effect could be affected by the material texture. The key parameters of two Ti64 $\alpha$ alloys with a loading direction 45$^{\circ}$ to each other were compared, however, there was no difference between them at a macroscopic level. The ratio of ‘hard’ to ‘soft’ grains in different macroscopic directions could lead to the activation of different slip systems during deformation. The effect of texture on time dependent plasticity should be studied in the context of activation of slip systems.

\section*{Acknowledgements}
The authors acknowledge funding from the EPSRC through the HexMat programme grant (EP/K034332/1) and the Diamond Light Source for beam time under experiment EE17222. We are grateful for use of characterisation facilities within the David Cockayne Centre for Electron Microscopy, Department of Materials, University of Oxford, which has benefitted from financial support provided by the Henry Royce Institute (Grant ref EP/R010145/1). YX expresses gratitude to the financial support of China Scholarship Council (CSC) and ET acknowledges EPSRC for support through Fellowship grant (EP/N007239/1).

\appendix

\section{Sensitivity of the fitted parameters}
\label{sec:app}
\renewcommand{\thetable}{A\arabic{table}}
\setcounter{table}{0}

The sensitivity of the three fitted parameters ($\Delta{V}$, $\Delta{F}$ and $\sigma_c$) to the change of $\rho$ is studied here. The $\rho$ was changed by $\pm$40\%, following previous work~\cite{XIONG2021116937}. Same fitting method was utilised to find the best fitted $\Delta{V}$, $\Delta{F}$ and $\sigma_c$ of the Ti64 $\alpha$ samples. The results are shown in Table~\ref{table:A1}. It can be found that the three key parameters were not very sensitive to the changing $\rho$, best fitted parameters were still within the confidence intervals for the three parameters ($\pm0.2\times10^{-20}$J for $\Delta{F}$, $\pm1b^3$ for $\Delta{V}$ and $\pm5$MPa for $\sigma_c$).

\begin{table}[hbt!]
\caption{Sensitivity of the three fitted parameters ($\Delta{V}$, $\Delta{F}$ and $\sigma_c$) to the change of $\rho$ by $\pm$40\%. Using the Ti64 $\alpha$ samples as examples.}
\begin{tabular}{|c|c|c|c|c|}
\hline
                    & Change in $\rho$ & $\Delta{V}(b^3)$ & $\Delta{F}(10^{-20}\textrm{J})$ & $\sigma_c(\textrm{MPa})$ \\ \hline
\multirow{3}{*}{RT} &          $\rho$        &          11.71        &               10.28         &         871        \\ \cline{2-5} 
  &  $\rho+40\%$ &     11.86          &         10.42         &          873                   \\ \cline{2-5} 
                    &        $\rho-40\%$          &         11.82         & 10.11                       &         868        \\ \hline
\multirow{3}{*}{75$^{\circ}$C}   &       $\rho$          &       18.33           &            12.42            &          658       \\ \cline{2-5} 
                    &         $\rho+40\%$         &          18.02        & 12.58          &        656         \\ \cline{2-5} 
                    &        $\rho-40\%$          &         17.71         &     12.27                   &        655         \\ \hline
\multirow{3}{*}{145$^{\circ}$C}   &         $\rho$          &      34.30            &            15.62            &          562       \\ \cline{2-5} 
      &          $\rho+40\%$        &          34.87        &       15.74                 &         564        \\ \cline{2-5} 
                    &        $\rho-40\%$          &         34.50       &     15.46             &        558         \\ \hline
\multirow{3}{*}{250$^{\circ}$C}   &   $\rho$   &    85.05        &      21.02            &           560                             \\ \cline{2-5} 
                    &         $\rho+40\%$         &         84.99         &      21.17                  &      567           \\ \cline{2-5} 
                    &       $\rho-40\%$           &          86.27        &       20.90                 &         560        \\ \hline
\end{tabular}
\label{table:A1}
\end{table}

\bibliographystyle{elsarticle-num}
\bibliography{bib}

\begin{thebibliography}{10}
\expandafter\ifx\csname url\endcsname\relax
  \def\url#1{\texttt{#1}}\fi
\expandafter\ifx\csname urlprefix\endcsname\relax\def\urlprefix{URL }\fi
\expandafter\ifx\csname href\endcsname\relax
  \def\href#1#2{#2} \def\path#1{#1}\fi

\bibitem{BACHE2003}
M.~Bache,
  \href{http://www.sciencedirect.com/science/article/pii/S0142112303001452}{A
  review of dwell sensitive fatigue in titanium alloys: the role of
  microstructure, texture and operating conditions}, International Journal of
  Fatigue 25~(9) (2003) 1079 -- 1087, international Conference on Fatigue
  Damage of Structural Materials IV.
\newblock \href {https://doi.org/10.1016/S0142-1123(03)00145-2}
  {\path{doi:10.1016/S0142-1123(03)00145-2}}.
\newline\urlprefix\url{http://www.sciencedirect.com/science/article/pii/S0142112303001452}

\bibitem{DUNNE2008}
F.~P.~E. DUNNE, D.~RUGG,
  \href{https://onlinelibrary.wiley.com/doi/abs/10.1111/j.1460-2695.2008.01284.x}{On
  the mechanisms of fatigue facet nucleation in titanium alloys}, Fatigue \&
  Fracture of Engineering Materials \& Structures 31~(11) (2008) 949--958.
\newblock \href
  {http://arxiv.org/abs/https://onlinelibrary.wiley.com/doi/pdf/10.1111/j.1460-2695.2008.01284.x}
  {\path{arXiv:https://onlinelibrary.wiley.com/doi/pdf/10.1111/j.1460-2695.2008.01284.x}},
  \href {https://doi.org/10.1111/j.1460-2695.2008.01284.x}
  {\path{doi:10.1111/j.1460-2695.2008.01284.x}}.
\newline\urlprefix\url{https://onlinelibrary.wiley.com/doi/abs/10.1111/j.1460-2695.2008.01284.x}

\bibitem{ZHENG201743}
Z.~Zheng, D.~S. Balint, F.~P. Dunne,
  \href{http://www.sciencedirect.com/science/article/pii/S1359645417300319}{Investigation
  of slip transfer across hcp grain boundaries with application to cold dwell
  facet fatigue}, Acta Materialia 127 (2017) 43 -- 53.
\newblock \href {https://doi.org/https://doi.org/10.1016/j.actamat.2017.01.021}
  {\path{doi:https://doi.org/10.1016/j.actamat.2017.01.021}}.
\newline\urlprefix\url{http://www.sciencedirect.com/science/article/pii/S1359645417300319}

\bibitem{NEERAJ2000}
T.~Neeraj, D.-H. Hou, G.~Daehn, M.~Mills,
  \href{http://www.sciencedirect.com/science/article/pii/S1359645499004267}{Phenomenological
  and microstructural analysis of room temperature creep in titanium alloys},
  Acta Materialia 48~(6) (2000) 1225 -- 1238.
\newblock \href {https://doi.org/https://doi.org/10.1016/S1359-6454(99)00426-7}
  {\path{doi:https://doi.org/10.1016/S1359-6454(99)00426-7}}.
\newline\urlprefix\url{http://www.sciencedirect.com/science/article/pii/S1359645499004267}

\bibitem{CONRAD1981}
H.~Conrad,
  \href{http://www.sciencedirect.com/science/article/pii/0079642581900013}{Effect
  of interstitial solutes on the strength and ductility of titanium}, Progress
  in Materials Science 26~(2) (1981) 123 -- 403.
\newblock \href {https://doi.org/https://doi.org/10.1016/0079-6425(81)90001-3}
  {\path{doi:https://doi.org/10.1016/0079-6425(81)90001-3}}.
\newline\urlprefix\url{http://www.sciencedirect.com/science/article/pii/0079642581900013}

\bibitem{CONRAD2011}
H.~Conrad, Thermally activated deformation of titanium below 0.4 tm, Canadian
  Journal of Physics 45 (2011) 581--590.
\newblock \href {https://doi.org/10.1139/p67-050} {\path{doi:10.1139/p67-050}}.

\bibitem{LEI202177}
Z.~Lei, P.~Gao, X.~Wang, M.~Zhan, H.~Li,
  \href{https://www.sciencedirect.com/science/article/pii/S1005030221001766}{Analysis
  of anisotropy mechanism in the mechanical property of titanium alloy tube
  formed through hot flow forming}, Journal of Materials Science \& Technology
  86 (2021) 77--90.
\newblock \href {https://doi.org/https://doi.org/10.1016/j.jmst.2021.01.038}
  {\path{doi:https://doi.org/10.1016/j.jmst.2021.01.038}}.
\newline\urlprefix\url{https://www.sciencedirect.com/science/article/pii/S1005030221001766}

\bibitem{ZHANG2021265}
Z.~Zhang, J.~Fan, R.~Li, H.~Kou, Z.~Chen, Q.~Wang, H.~Zhang, J.~Wang, Q.~Gao,
  J.~Li,
  \href{https://www.sciencedirect.com/science/article/pii/S1005030220308859}{Orientation
  dependent behavior of tensile-creep deformation of hot rolled ti65 titanium
  alloy sheet}, Journal of Materials Science \& Technology 75 (2021) 265--275.
\newblock \href {https://doi.org/https://doi.org/10.1016/j.jmst.2020.10.021}
  {\path{doi:https://doi.org/10.1016/j.jmst.2020.10.021}}.
\newline\urlprefix\url{https://www.sciencedirect.com/science/article/pii/S1005030220308859}

\bibitem{DUNNE20071061}
F.~Dunne, D.~Rugg, A.~Walker,
  \href{http://www.sciencedirect.com/science/article/pii/S0749641906001641}{Lengthscale-dependent,
  elastically anisotropic, physically-based hcp crystal plasticity: Application
  to cold-dwell fatigue in ti alloys}, International Journal of Plasticity
  23~(6) (2007) 1061 -- 1083.
\newblock \href {https://doi.org/10.1016/j.ijplas.2006.10.013}
  {\path{doi:10.1016/j.ijplas.2006.10.013}}.
\newline\urlprefix\url{http://www.sciencedirect.com/science/article/pii/S0749641906001641}

\bibitem{HASIJA2003}
V.~Hasija, S.~Ghosh, M.~J. Mills, D.~S. Joseph,
  \href{http://www.sciencedirect.com/science/article/pii/S1359645403002891}{Deformation
  and creep modeling in polycrystalline ti–6al alloys}, Acta Materialia
  51~(15) (2003) 4533 -- 4549.
\newblock \href {https://doi.org/10.1016/S1359-6454(03)00289-1}
  {\path{doi:10.1016/S1359-6454(03)00289-1}}.
\newline\urlprefix\url{http://www.sciencedirect.com/science/article/pii/S1359645403002891}

\bibitem{Qiu2014}
J.~Qiu, Y.~Ma, J.~Lei, Y.~Liu, A.~Huang, D.~Rugg, R.~Yang,
  \href{https://doi.org/10.1007/s11661-014-2541-5}{A comparative study on dwell
  fatigue of ti-6al-2sn-4zr-xmo (x=2 to 6) alloys on a
  microstructure-normalized basis}, Metallurgical and Materials Transactions A
  45~(13) (2014) 6075--6087.
\newblock \href {https://doi.org/10.1007/s11661-014-2541-5}
  {\path{doi:10.1007/s11661-014-2541-5}}.
\newline\urlprefix\url{https://doi.org/10.1007/s11661-014-2541-5}

\bibitem{TITANIUM}
G.~L{\"u}tjering, J.~Williams,
  \href{https://books.google.co.uk/books?id=GwI9ul\_wAegC}{Titanium},
  Engineering materials and processes, Springer, 2003.
\newline\urlprefix\url{https://books.google.co.uk/books?id=GwI9ul\_wAegC}

\bibitem{ZHANG2015}
Z.~Zhang, M.~Cuddihy, F.~Dunne, On rate-dependent polycrystal deformation: the
  temperature sensitivity of cold dwell fatigue, Proceedings of the Royal
  Society A: Mathematical, Physical and Engineering Science 471 (2015)
  20150214.
\newblock \href {https://doi.org/10.1098/rspa.2015.0214}
  {\path{doi:10.1098/rspa.2015.0214}}.

\bibitem{TANG2021116468}
Y.~T. Tang, N.~D’Souza, B.~Roebuck, P.~Karamched, C.~Panwisawas, D.~M.
  Collins,
  \href{https://www.sciencedirect.com/science/article/pii/S1359645420308855}{Ultra-high
  temperature deformation in a single crystal superalloy: Mesoscale process
  simulation and micromechanisms}, Acta Materialia 203 (2021) 116468.
\newblock \href {https://doi.org/https://doi.org/10.1016/j.actamat.2020.11.010}
  {\path{doi:https://doi.org/10.1016/j.actamat.2020.11.010}}.
\newline\urlprefix\url{https://www.sciencedirect.com/science/article/pii/S1359645420308855}

\bibitem{Roebuck2001}
B.~Roebuck, D.~Cox, R.~Reed, The temperature dependence of $\gamma'$ volume
  fraction in a ni-based single crystal superalloy from resistivity
  measurements, Scripta Materialia 44 (2001) 917--921.

\bibitem{SULZER2018}
S.~Sulzer, E.~Alabort, A.~Németh, B.~Roebuck, R.~Reed, On the rapid assessment
  of mechanical behavior of a prototype nickel-based superalloy using
  small-scale testing, Metallurgical and Materials Transactions A 49 (2018)
  4214–4235.
\newblock \href {https://doi.org/10.1007/s11661-018-4673-5}
  {\path{doi:10.1007/s11661-018-4673-5}}.

\bibitem{XIONG2020}
Y.~Xiong, P.~S. Karamched, C.-T. Nguyen, D.~M. Collins, C.~M. Magazzeni,
  E.~Tarleton, A.~J. Wilkinson,
  \href{http://www.sciencedirect.com/science/article/pii/S1359645420306078}{Cold
  creep of titanium: Analysis of stress relaxation using synchrotron
  diffraction and crystal plasticity simulations}, Acta Materialia 199 (2020)
  561 -- 577.
\newblock \href {https://doi.org/https://doi.org/10.1016/j.actamat.2020.08.010}
  {\path{doi:https://doi.org/10.1016/j.actamat.2020.08.010}}.
\newline\urlprefix\url{http://www.sciencedirect.com/science/article/pii/S1359645420306078}

\bibitem{creep_analysis_constitutive}
H.~ALTENBACH, K.~NAUMENKO, Y.~GORASH,
  \href{https://doi.org/10.1142/S0217979208050589}{Creep analysis for a wide
  stress range based on stress relaxation experiments}, International Journal
  of Modern Physics B 22~(31n32) (2008) 5413--5418.
\newblock \href
  {http://arxiv.org/abs/https://doi.org/10.1142/S0217979208050589}
  {\path{arXiv:https://doi.org/10.1142/S0217979208050589}}, \href
  {https://doi.org/10.1142/S0217979208050589}
  {\path{doi:10.1142/S0217979208050589}}.
\newline\urlprefix\url{https://doi.org/10.1142/S0217979208050589}

\bibitem{Kalyanasundaram2016}
V.~Kalyanasundaram, S.~R. Holdsworth,
  \href{https://doi.org/10.1007/s12666-015-0776-5}{Prediction of forward creep
  behaviour from stress relaxation data for a 10 {\%} cr steel at
  600 {\textdegree}c}, Transactions of the Indian Institute of Metals 69~(2)
  (2016) 573--578.
\newblock \href {https://doi.org/10.1007/s12666-015-0776-5}
  {\path{doi:10.1007/s12666-015-0776-5}}.
\newline\urlprefix\url{https://doi.org/10.1007/s12666-015-0776-5}

\bibitem{WANG2016656}
Y.~Wang, M.~Spindler, C.~Truman, D.~Smith,
  \href{https://www.sciencedirect.com/science/article/pii/S0264127516301162}{Critical
  analysis of the prediction of stress relaxation from forward creep of type
  316h austenitic stainless steel}, Materials \& Design 95 (2016) 656--668.
\newblock \href {https://doi.org/https://doi.org/10.1016/j.matdes.2016.01.118}
  {\path{doi:https://doi.org/10.1016/j.matdes.2016.01.118}}.
\newline\urlprefix\url{https://www.sciencedirect.com/science/article/pii/S0264127516301162}

\bibitem{ZHANG2016Intrinsic}
Z.~Zhang, T.-S. Jun, T.~B. Britton, F.~P. Dunne,
  \href{http://www.sciencedirect.com/science/article/pii/S1359645416305523}{Intrinsic
  anisotropy of strain rate sensitivity in single crystal alpha titanium}, Acta
  Materialia 118 (2016) 317 -- 330.
\newblock \href {https://doi.org/https://doi.org/10.1016/j.actamat.2016.07.044}
  {\path{doi:https://doi.org/10.1016/j.actamat.2016.07.044}}.
\newline\urlprefix\url{http://www.sciencedirect.com/science/article/pii/S1359645416305523}

\bibitem{ZHANG2017199}
Z.~Zhang, F.~P. Dunne,
  \href{https://www.sciencedirect.com/science/article/pii/S0022509616306858}{Microstructural
  heterogeneity in rate-dependent plasticity of multiphase titanium alloys},
  Journal of the Mechanics and Physics of Solids 103 (2017) 199--220.
\newblock \href {https://doi.org/https://doi.org/10.1016/j.jmps.2017.03.012}
  {\path{doi:https://doi.org/10.1016/j.jmps.2017.03.012}}.
\newline\urlprefix\url{https://www.sciencedirect.com/science/article/pii/S0022509616306858}

\bibitem{ASHTON2017377}
P.~J. Ashton, T.-S. Jun, Z.~Zhang, T.~B. Britton, A.~M. Harte, S.~B. Leen,
  F.~P. Dunne,
  \href{https://www.sciencedirect.com/science/article/pii/S014211231730138X}{The
  effect of the beta phase on the micromechanical response of dual-phase
  titanium alloys}, International Journal of Fatigue 100 (2017) 377--387.
\newblock \href
  {https://doi.org/https://doi.org/10.1016/j.ijfatigue.2017.03.020}
  {\path{doi:https://doi.org/10.1016/j.ijfatigue.2017.03.020}}.
\newline\urlprefix\url{https://www.sciencedirect.com/science/article/pii/S014211231730138X}

\bibitem{XIONG2021116937}
Y.~Xiong, P.~S. Karamched, C.-T. Nguyen, D.~M. Collins, N.~Grilli, C.~M.
  Magazzeni, E.~Tarleton, A.~J. Wilkinson,
  \href{https://www.sciencedirect.com/science/article/pii/S1359645421003177}{An
  in-situ synchrotron diffraction study of stress relaxation in titanium:
  Effect of temperature and oxygen on cold dwell fatigue}, Acta Materialia
  (2021) 116937\href
  {https://doi.org/https://doi.org/10.1016/j.actamat.2021.116937}
  {\path{doi:https://doi.org/10.1016/j.actamat.2021.116937}}.
\newline\urlprefix\url{https://www.sciencedirect.com/science/article/pii/S1359645421003177}

\bibitem{CUDDIHY2017}
M.~Cuddihy, A.~Stapleton, S.~Williams, F.~Dunne,
  \href{http://www.sciencedirect.com/science/article/pii/S0142112316304054}{On
  cold dwell facet fatigue in titanium alloy aero-engine components},
  International Journal of Fatigue 97 (2017) 177 -- 189.
\newblock \href
  {https://doi.org/https://doi.org/10.1016/j.ijfatigue.2016.11.034}
  {\path{doi:https://doi.org/10.1016/j.ijfatigue.2016.11.034}}.
\newline\urlprefix\url{http://www.sciencedirect.com/science/article/pii/S0142112316304054}

\bibitem{STRINGER1960}
J.~Stringer,
  \href{http://www.sciencedirect.com/science/article/pii/000161606090170X}{The
  oxidation of titanium in oxygen at high temperatures}, Acta Metallurgica
  8~(11) (1960) 758 -- 766.
\newblock \href {https://doi.org/https://doi.org/10.1016/0001-6160(60)90170-X}
  {\path{doi:https://doi.org/10.1016/0001-6160(60)90170-X}}.
\newline\urlprefix\url{http://www.sciencedirect.com/science/article/pii/000161606090170X}

\bibitem{boyle2013stress}
J.~Boyle, J.~Spence,
  \href{https://books.google.co.uk/books?id=3b38BAAAQBAJ}{Stress Analysis for
  Creep}, Elsevier Science, 2013.
\newline\urlprefix\url{https://books.google.co.uk/books?id=3b38BAAAQBAJ}

\bibitem{TANAKA2016}
M.~Tanaka, K.~Higashida,
  \href{https://doi.org/10.1080/14786435.2016.1183828}{Temperature dependence
  of effective stress in severely deformed ultralow-carbon steel},
  Philosophical Magazine 96~(19) (2016) 1979--1992.
\newblock \href
  {http://arxiv.org/abs/https://doi.org/10.1080/14786435.2016.1183828}
  {\path{arXiv:https://doi.org/10.1080/14786435.2016.1183828}}, \href
  {https://doi.org/10.1080/14786435.2016.1183828}
  {\path{doi:10.1080/14786435.2016.1183828}}.
\newline\urlprefix\url{https://doi.org/10.1080/14786435.2016.1183828}

\bibitem{Yu2015}
Q.~Yu, L.~Qi, T.~Tsuru, R.~Traylor, D.~Rugg, J.~W. Morris, M.~Asta, D.~C.
  Chrzan, A.~M. Minor,
  \href{https://science.sciencemag.org/content/347/6222/635}{Origin of dramatic
  oxygen solute strengthening effect in titanium}, Science 347~(6222) (2015)
  635--639.
\newblock \href
  {http://arxiv.org/abs/https://science.sciencemag.org/content/347/6222/635.full.pdf}
  {\path{arXiv:https://science.sciencemag.org/content/347/6222/635.full.pdf}},
  \href {https://doi.org/10.1126/science.1260485}
  {\path{doi:10.1126/science.1260485}}.
\newline\urlprefix\url{https://science.sciencemag.org/content/347/6222/635}

\bibitem{BRITTON2015}
T.~B. Britton, F.~P.~E. Dunne, A.~J. Wilkinson,
  \href{https://royalsocietypublishing.org/doi/abs/10.1098/rspa.2014.0881}{On
  the mechanistic basis of deformation at the microscale in hexagonal
  close-packed metals}, Proceedings of the Royal Society A: Mathematical,
  Physical and Engineering Sciences 471~(2178) (2015) 20140881.
\newblock \href
  {http://arxiv.org/abs/https://royalsocietypublishing.org/doi/pdf/10.1098/rspa.2014.0881}
  {\path{arXiv:https://royalsocietypublishing.org/doi/pdf/10.1098/rspa.2014.0881}},
  \href {https://doi.org/10.1098/rspa.2014.0881}
  {\path{doi:10.1098/rspa.2014.0881}}.
\newline\urlprefix\url{https://royalsocietypublishing.org/doi/abs/10.1098/rspa.2014.0881}

\bibitem{CORREA2015180}
D.~R.~N. Correa, F.~B. Vicente, R.~O. Araújo, M.~L. Lourenço, P.~A.~B.
  Kuroda, M.~A.~R. Buzalaf, C.~R. Grandini,
  \href{https://www.sciencedirect.com/science/article/pii/S2238785415000320}{Effect
  of the substitutional elements on the microstructure of the ti-15mo-zr and
  ti-15zr-mo systems alloys}, Journal of Materials Research and Technology
  4~(2) (2015) 180--185.
\newblock \href {https://doi.org/https://doi.org/10.1016/j.jmrt.2015.02.007}
  {\path{doi:https://doi.org/10.1016/j.jmrt.2015.02.007}}.
\newline\urlprefix\url{https://www.sciencedirect.com/science/article/pii/S2238785415000320}

\bibitem{Sarkar2015}
A.~Sarkar, J.~K. Chakravartty,
  \href{https://doi.org/10.1007/s11661-015-3153-4}{Activation volume and
  density of mobile dislocations in plastically deforming
  zr-1pctsn-1pctnb-0.1pctfe alloy}, Metallurgical and Materials Transactions A
  46~(12) (2015) 5638--5643.
\newblock \href {https://doi.org/10.1007/s11661-015-3153-4}
  {\path{doi:10.1007/s11661-015-3153-4}}.
\newline\urlprefix\url{https://doi.org/10.1007/s11661-015-3153-4}

\bibitem{Fellah2019}
M.~Fellah, N.~Hezil, M.~Abdul~Samad, R.~Djellabi, A.~Montagne, A.~Mejias,
  S.~Kossman, A.~Iost, A.~Purnama, A.~Obrosov, S.~Weiss,
  \href{https://doi.org/10.1007/s11665-019-04348-w}{Effect of molybdenum
  content on structural, mechanical, and tribological properties of hot
  isostatically pressed $\beta$-type titanium alloys for orthopedic
  applications}, Journal of Materials Engineering and Performance 28~(10)
  (2019) 5988--5999.
\newblock \href {https://doi.org/10.1007/s11665-019-04348-w}
  {\path{doi:10.1007/s11665-019-04348-w}}.
\newline\urlprefix\url{https://doi.org/10.1007/s11665-019-04348-w}

\bibitem{WANG20062715}
Y.~Wang, A.~Hamza, E.~Ma,
  \href{https://www.sciencedirect.com/science/article/pii/S1359645406001364}{Temperature-dependent
  strain rate sensitivity and activation volume of nanocrystalline ni}, Acta
  Materialia 54~(10) (2006) 2715--2726.
\newblock \href {https://doi.org/https://doi.org/10.1016/j.actamat.2006.02.013}
  {\path{doi:https://doi.org/10.1016/j.actamat.2006.02.013}}.
\newline\urlprefix\url{https://www.sciencedirect.com/science/article/pii/S1359645406001364}

\bibitem{CHENG20034505}
S.~Cheng, J.~Spencer, W.~Milligan,
  \href{https://www.sciencedirect.com/science/article/pii/S1359645403002866}{Strength
  and tension/compression asymmetry in nanostructured and ultrafine-grain
  metals}, Acta Materialia 51~(15) (2003) 4505--4518.
\newblock \href {https://doi.org/https://doi.org/10.1016/S1359-6454(03)00286-6}
  {\path{doi:https://doi.org/10.1016/S1359-6454(03)00286-6}}.
\newline\urlprefix\url{https://www.sciencedirect.com/science/article/pii/S1359645403002866}

\bibitem{Evans_and_Rawlings}
A.~G. Evans, R.~D. Rawlings,
  \href{https://onlinelibrary.wiley.com/doi/abs/10.1002/pssb.19690340102}{The
  thermally activated deformation of crystalline materials}, physica status
  solidi (b) 34~(1) (1969) 9--31.
\newblock \href
  {http://arxiv.org/abs/https://onlinelibrary.wiley.com/doi/pdf/10.1002/pssb.19690340102}
  {\path{arXiv:https://onlinelibrary.wiley.com/doi/pdf/10.1002/pssb.19690340102}},
  \href {https://doi.org/https://doi.org/10.1002/pssb.19690340102}
  {\path{doi:https://doi.org/10.1002/pssb.19690340102}}.
\newline\urlprefix\url{https://onlinelibrary.wiley.com/doi/abs/10.1002/pssb.19690340102}

\bibitem{JUN2016}
T.-S. Jun, Z.~Zhang, G.~Sernicola, F.~P. Dunne, T.~B. Britton,
  \href{http://www.sciencedirect.com/science/article/pii/S1359645416300544}{Local
  strain rate sensitivity of single $\alpha$ phase within a dual-phase ti
  alloy}, Acta Materialia 107 (2016) 298 -- 309.
\newblock \href {https://doi.org/https://doi.org/10.1016/j.actamat.2016.01.057}
  {\path{doi:https://doi.org/10.1016/j.actamat.2016.01.057}}.
\newline\urlprefix\url{http://www.sciencedirect.com/science/article/pii/S1359645416300544}

\bibitem{JUN2016NANO}
T.-S. Jun, D.~E. Armstrong, T.~B. Britton,
  \href{http://www.sciencedirect.com/science/article/pii/S0925838816304133}{A
  nanoindentation investigation of local strain rate sensitivity in dual-phase
  ti alloys}, Journal of Alloys and Compounds 672 (2016) 282 -- 291.
\newblock \href {https://doi.org/https://doi.org/10.1016/j.jallcom.2016.02.146}
  {\path{doi:https://doi.org/10.1016/j.jallcom.2016.02.146}}.
\newline\urlprefix\url{http://www.sciencedirect.com/science/article/pii/S0925838816304133}

\end{thebibliography}

\end{document}